\documentstyle[12pt,prl,aps]{revtex}

\draft
\begin{document}
\title{ 
Quantum Monte Carlo Calculations for a large number of bosons in
a harmonic trap}
\author{Werner Krauth}
\address{
CNRS-Laboratoire de Physique Statistique de l'ENS\\
24, rue Lhomond; F-75231 Paris Cedex 05; France\\
e-mail: krauth@physique.ens.fr\\
}
\date{June, 1996}
\maketitle
\begin{abstract}
In this paper, I present a precise Quantum Monte Carlo calculation
at finite temperature for a very large number (many thousands) 
of bosons in a harmonic trap, which may be anisotropic.
The calculation  applies directly to the recent experiments of 
Bose-Einstein condensation of atomic vapors in magnetic traps.
I show that the critical temperature of the system decreases 
with the interaction. I also present profiles for the 
overall density and the one of condensed particles, and obtain 
excellent agreement with solutions of the 
Gross-Pitaevskii equation. 

\end{abstract}
\newpage
The  achievement of Bose-Einstein condensation (BEC) in dilute 
atomic vapors of $^{87}$Rb \cite{Cornell} and $^{23}$Na \cite{Ketterle}
has created extraordinary 
experimental and theoretical activity. 
The experiments pose fundamental 
theoretical questions, ranging from a clear understanding of the 
dynamics of BEC in a finite sample \cite{Griffin} to the behavior 
under time-dependent perturbations.

Besides these tantalizing and very controversial questions about the 
dynamics  of the quantum coherence, there are many questions
about the {\em statics} (thermodynamics) of the BE-condensate in a harmonic
trap. 
In this paper, I show that we can obtain the static properties from 
Path-Integral Quantum Monte Carlo (QMC) calculations 
which suffer from no systematic uncertainties other
than the choice of the interatomic (pseudo-)potential. 
I will present results for $10000$ particles in isotropic
and anisotropic traps around the BEC transition point, {\em i.e.} 
at the temperatures and 
particle number of current experimental interest. 

The hamiltonian of the system is given by
\begin{equation}
H= \sum_{i=1}^N p_i^2/2 m + \sum_{i,j} V(r_{ij}) +
 \frac{m}{2} (\omega_x^2 x^2 + \omega_y^2 y^2 + \omega_z^2 z^2)
\label{hamilton}
\end{equation}
where $V$ is the interatomic potential \cite{Krauss}. This potential has 
many bound states, which, in the experiment, 
are inaccessible
due to the kinematics of two-particle collisions.
In thermodynamic calculations, as the one presented here,  
it is necessary to work with a pseudo-potential. I simply adopt
a hard-core term of radius $a_0$
- as others have done before. More refined choices can easily be implemented.

The physical parameters - and the units - are 
as in ref. \cite{Dalfovo}: we consider two  harmonic traps (isotropic and
anisotropic), and units in which
all frequencies are equal to $1$ in the isotropic case, 
and $\omega_x = \omega_y =1$, 
$\omega_z = \sqrt{8}$ in the anisotropic case. The anisotropy 
seems to be close to the experimental situation. The value $a_0$ for the
hard-core radius is taken to be $a_0 = 0.0043$ in the above units 
\cite{Dalfovo}.

To situate the computations of the interacting case, it is useful
to remember the basic result for non-interacting  bosons (in our 
isotropic potential, and with the above units) \cite{deGroot} \cite{Bagnato}: 
for large $N$,  the
transition temperature $T_{c}$ scales as $T_{c} \sim 
\tilde{T_{c}} N^{1/3}$ with $\tilde{T_{c}}  \sim 0.94$.
For $N=10.000$ particles this gives 
$\beta_c \sim 0.05$. Let us mention in passing that at this value
of $N$ there are no more detectable differences between 
the canonical and the grand-canonical formalism \cite{footcanonic}.

The partition function $Z$, the trace over the 
density matrix,  satisfies the usual convolution equation:
\begin{equation}
Z= \frac{1}{N !}\sum_P  \int d R \rho(R,R^P,\beta)=\frac{1}{N !}\sum_P
   \int \ldots \int d R d R_2 \ldots d R_{M} \rho(R,R_2,\tau) \ldots
\rho(R_{M},R^P,\tau)
\label{convolution}
\end{equation}
Here $\tau = \beta/M$; $R$ is the 
$3 N$-dimensional vector of the particle
coordinates $R=(r_1,r_2,\ldots,r_N)$, and $R^P$ denotes the vector 
with permuted particle labels:
$R^P=(r_{P(1)},r_{P(2)},\ldots,r_{P(N)})$. 

In the limit  $M\rightarrow \infty$, the exact convolution 
equation eq.~(\ref{convolution}) reduces to the Feynman-Kac formula
({\em cf} \cite{Feynman}), as a consequence of the Trotter break-up
$\exp[\tau (H_1 + H_2)] \sim \exp[\tau H_1] \exp[\tau H_2]$ for 
$\tau \rightarrow 0$.

However,  it is the basic technical lesson of past QMC work 
\cite{Barker} \cite{Ceperley}
that eq.~(\ref{convolution}) should only be taken to intermediate temperatures
$\tau$ such that the approximation of the complete density 
matrix $\rho(R,R',\tau)$
by a product of pair terms becomes accurate
\begin{equation}
\rho(R,R',\tau)  \simeq \prod_i\rho_1(r_i,r'_i,\tau) \prod_{i<j} 
\frac{\rho_2(r_i,r_j,r_i',r_j',\tau)}
{\rho_1(r_i,r'_i,\tau) \rho_1(r_j,r'_j,\tau)}
\label{pairansatz}
\end{equation}
The one-particle density matrices $\rho_1(r_i,r_i',\tau)$ in 
eq.~(\ref{pairansatz})
describe a particle in the harmonic potential: an exact solution 
is known ({\em cf.}, {\em e.g.} \cite{Feynman}), 
and can easily be sampled. This product of one-particle density matrices
is used as {\em a priori} probability density in the sense of 
ref. \cite{Ceperley}. A very efficient method is used to generate a
sequence of configurations, and to sample permutations. 
The second part in eq. (\ref{pairansatz})
is equal to $1$  for $V=0$. It plays the role of a 
correction term due to two-body interaction. The Metropolis 
algorithm is used to incorporate this term, and to generate configurations
which are distributed according to eq.~(\ref{convolution}).

To obtain the correction factor we need the two-particle density 
matrix. Fortunately, for the special case of the harmonic potential,
the two-particle hamiltonian - and therefore the density matrix -  
decouples into a
center-of-mass term (which is unaffected by the interaction) and a
term describing the relative motion. The latter is given by:
\begin{equation}
   H_{rel} =  p^2/2 \mu + V(r) +
 \frac{\mu}{2} (\omega_x^2 x^2 + \omega_y^2 y^2 + \omega_z^2 z^2)
\label{relative}
\end{equation}
where $\mu = m/2$ is the reduced mass.

The eigenfunctions of the pure hard-sphere potential
(eq.~(\ref{relative})
without the terms in $\omega$) 
can be obtained exactly (\cite{Larsen}). 
Adding an isotropic trap potential, the same may still be possible.
However, since I am interested in treating the 
general {\em anisotropic} trap, 
I have obtained the relative-motion density matrix 
from the exact solution of the 
hard-core potential density matrix $\rho_{hc}$,
into which I incorporate the harmonic term  
{\em via} the Trotter break-up. 
\begin{equation}
  \rho_{2,rel}(r,r',\tau)\simeq X(r) \times
 \rho_{hc}(r,r') \times X(r')
\label{trotterpair}
\end{equation}
with $X(r)= \exp[-\tau\mu
(\omega_x^2 x^2 + \omega_y^2 y^2 + \omega_z^2 z^2)/4] $.
The $\tau$ finally retained in the simulation has to be chosen to accommodate
both eq.~(\ref{pairansatz}) and eq.~(\ref{trotterpair}). Extensive tests
have convinced me that for the density attained, 
a value of $\tau=0.01$ is appropriate. This value
is about $6$ {\em orders of magnitude} larger than the one which would have to
be used with the simple Trotter break-up. 

For the value of $\tau=0.01$ which was found sufficient, we can explicitly 
compute the
effective {\em range} of the interaction, beyond which the correction factor in 
eq.~(\ref{pairansatz}) is practically one. This range turns out to be about 
$0.2$. It is evident that one may introduce a $3-$dimensional grid, with
the grid size larger than the interaction range. 
At any time,  particles are assigned to boxes formed by the grid. 
In evaluating the pair density 
matrix eq.~(\ref{pairansatz}), an efficient algorithm can be set up
to compute the correction term only for close-by pairs. 
As a result, the program behaves gracefully as $N$ is increased,
and the actual limit of the calculation
is rather given by the memory requirements than by CPU considerations.

The program has been checked very carefully. In the non-interacting case, 
I am able to reproduce all the exactly known results. For
an isotropic trap, and in the absence of interactions, {\em e. g.} 
I have reproduced
the analytically known results for the condensate 
fraction $N_0/N$, which is plotted in fig.~1. I also plot the 
same quantity with interactions:
the condensate fraction clearly decreases with respect to the 
noninteracting case. 
From this I conclude that $T_c$ decreases. Notice that the 
interaction influences the condensate fraction  much more  than 
the finite-size effects for the non-interacting gas, which are also shown.

On the left side of  fig.~2, I show the corresponding density profiles 
$\rho(x) = \int \int dy dz \rho(x,y,z)$ 
of the particles.  The center density increases
sharply as $T$ is lowered. This is the hallmark of real-space BEC in the 
confining potential. 

We pause for a moment to discuss BEC in the path-integral framework:
below $T_{c}$, particles have a finite probability to belong
to extended permutation cycles of length $l$. 
Notice that in the confined geometry the off-diagonal 
one-particle density matrix
at large separations trivially vanishes, instead of going to a 
value proportional to $N_0$ (as in the translational
invariant system). In the present simulation, I instead obtain the 
number of condensed particles from the permutations of the system:
the maximum length of 
$l$ which has non-zero probability is equal to $N_0$.  
For non-interacting particles in the 
limit $T\rightarrow 0$, the probability to belong to extended
permutation cycles is independent of $l$ (having all 
particles in the condensate does therefore not mean that they are all on the
same cycle). At any temperature, noninteracting bosons which are on the 
different permutation cycles are statistically independent \cite{Ceperley} 
- a property which is used
in the  QMC code to generate the {\em a priori} probability. 
The spatial distribution of particles in a cycle of length 
$l$ is given by the diagonal density matrix at temperature $l  \beta$
\cite{Feynman}:
\begin{equation}
\rho(x,x,l\beta) \sim \exp( -m \omega x^2 \tanh \frac{\omega}{2} \beta l)
\label{diagdens}
\end{equation}
Since $\beta_c \sim 1/N^{1/3}$, the particles
on long cycles with $l >> N^{1/3}$ (especially $l \sim N$)
are distributed according
to the lowest single-particle state $\Psi_0^2(x) \sim  \exp( -m \omega x^2)$.

As we introduce interactions, we have to give up the concept of condensation
into the lowest single-particle state.
However, we preserve the two other
essential features: below the transition, long permutation appear, 
and particles
on long permutation cycles are distributed identically: they populate the 
macroscopic quantum state.
We can thus gain access at the distribution of the condensed particles
by computing the density profiles as before, but restricted to particles
on cycles longer than some value $l_{min}$. I have verified explicitly
that for $l_{min} \stackrel{>}{\sim} 20$ the density profile does not
depend on this parameter, {\em i. e.} that the condensation concerns the 
ground state. The density profiles for the condensed particles is
given on the right side of fig.~(2);
the distributions are normalized to $1$ at any temperature. As we expect, 
the distribution of condensed particles is broader at low
temperature, where the  many
particles in the condensate strongly repel each other. 

Can we  obtain a more quantitative description of the condensate at 
finite temperature? To answer the question, I have 
computed the solution of the isotropic Gross-Pitaevskii (G-P) 
equation \cite{Gross}
\cite{Burnett} for the same value of $a_0$ as is used in 
the QMC-calculation, and for the number of particles corresponding
approximately to the condensate fraction, as obtained from fig.~1.
The results \cite{footsolution} are plotted
together with $\rho(x)$ for the 
condensed particles on the right side of fig.~2. The agreement is 
truly remarkable for all the three curves, especially since there
are no adjustable parameters. Even the number 
of condensed particles has been simply taken from fig.~1, without 
trying to optimize the fit. It can thus be said that, to a very high 
precision, the condensed particles are described by the G-P wavefunction.

The numerical results presented in fig.~2 suggest the following
quantitatively correct picture for the Bose-Einstein condensation 
in a trap at finite temperature:
below the (interaction-dependent) $T_c$, $N_0(T)$ 
particles are condensed. To a very high precision, these particles are 
distributed according to the appropriate G-P wavefunction. The non-condensed
particles are distributed as in the free case and are very much
spread out. There is very 
little mutual interaction between condensed and non-condensed particles:
on the one hand, the non-condensed particles are very dilute 
in the central region of the trap, where most of the particles present
are ``condensed'', on the other hand the G-P wavefunction disturbs
the non-condensed particles only on a small portion of their support.

The above picture allows us to understand the competition
of energy gain end entropy destruction which underlies the condensation 
into the macroscopic wavefunction. 
The balance of
entropy is the same for the non-interacting and interacting case, since
the non-condensed part is undisturbed by the interaction, and the 
condensed one has zero entropy.  
The energy of the G-P wavefunction is of course much higher
than in the absence of interactions. Condensation is therefore less 
favored, and we understand that the critical temperature must decrease
with the interaction, as shown in fig.~1. It will be very simple 
to perform a one-parameter (in $N_0$)  variational minimization 
of the free energy
which describes the non-condensed particles as free, and which uses
the energy of the G-P wavefunction. 

Rather than to proceed in this direction, I finally present some results
for the anisotropic trap, which is the case of direct interest to
experiments. In fig.~3, I present density plots for this case at 
different temperatures, and compare to the solution of 
the anisotropic G-P wavefunction with $N=10000$ \cite{Dalfovo}. 
As before, there is 
very close agreement between the two calculations. The calculation
also confirms that, in the limit of zero temperature, the 
number of condensed particles, $N_0$, seems to be extremely close to $N$.
This agrees with  recent work in which the interaction of the G-P 
groundstate with non-condensed particles was studied in the 
Bogolubov approximation \cite{Castin}. I have performed the same 
calculation as in fig.~1 for the isotropic case (the noninteracting
curve is easily computed, even for finite $N$ \cite{Bouchiat}), and the 
results are analogous. For example, at $\beta=0.06$, the noninteracting
gas in the anisotropic trap has $N_0/N=0.76$, but I only find a value
of $0.6$ for the interacting case.

Even though the   main impetus in this paper 
has been on the dependence  of the critical temperature
on the interaction, and on the comparison of the condensate wave function
with the Gross-Pitaevskii formalism, 
it should be evident that the QMC
calculation goes much beyond these results. We can obtain 
complete information on inter-particle correlation, compute
the complete thermodynamics of the system, {\em etc}. 
In particular, it is possible \cite{Sindzingre}
to obtain the nonclassical moment of inertia which has
attracted some attention lately \cite{Stringari}, 
and which will be of experimental relevance
as soon as rotating traps will become available. If the above picture is
correct, this moment (which basically follows the normal part of the 
gas) should depend on the interaction only through $N_0$.

I would also like to draw the reader's attention to the fact that
the very efficient algorithm is of general usefulness for the study 
of weakly interacting bosons. 
Other possible applications are 
bosons in $3$ dimensions without confining potential, but particularly
in  $2$ dimensions, where the effect of a small interaction 
will be even more important than in the case treated here, since the 
non-interacting gas has no phase transition at all.
To stimulate work in this area, 
as well as to facilitate direct comparison with experiences on trapped bosons, 
I will make available the FORTRAN code used in the present investigation.

\acknowledgments
I am grateful to F. Dalfovo for sending me data for the G-P density profiles
in the anisotropic trap, and to P. Patricio da Silva for explanations of his
finite-element program. C. Waigl participated in the initial stages of this
work. I thank C. Bouchiat for many discussions, and continuing encouragement,
and acknowledge very helpful discussions with 
Y. Castin,  J. Dalibard, R. Dum, P. Gr\"{u}ter, and F. Lalo\"e.

\noindent
\newpage
{\bf Figure Captions}

\begin{enumerate}
\item
Ratio of condensed particles $N_0/N$ {\em vs} reduced temperature 
$\tilde{T} = T N^{1/3}$ in 
an isotropic trap with $\omega=1$ for the noninteracting case in the 
thermodynamic limit  ($N\rightarrow \infty$, full line) and for 
$N=10000$ (dashed line), 
as well as for $10000$ interacting particles with $a_0= 0.0043$. The number
of condensed particles decreases with the interaction.
\item
Density profile {\em vs} $x$ for temperatures 
$\beta= 0.06 (\tilde{T}=0.78)$, $\beta= 0.07 (\tilde{T}=0.66)$, 
$\beta=0.12 (\tilde{T}=0.39)$ (left side, from below).
On the right side is plotted the density profile of the ``condensed''
particles for the same temperatures (from {\em above}, the curves are
normalized to one). 
These density profiles are in excellent agreement 
with the Gross-Pitaevskii solutions for the $N_0$ obtained from fig.~1 (dotted
lines). The values used are $N_0=2000$ for $\beta= 0.06$ (upper)
$N_0=4000$ for $\beta= 0.07$ (middle) $N_0=8000$ for $\beta= 0.12$ (lower).  

\item 
Density profile {\em vs} $x$ for an anisotropic trap with $N=10000$.
As in fig.~2, we plot the total density on the left, and the normalized density 
of condensed particles on the right. The temperatures are $\beta=0.06$ (dotted
line), and 
$\beta=0.16$ (full line).
The dashed curve is the  Gross-Pitaevskii solution for $N=10000$.
\end{enumerate}

\begin{references}
\bibitem{footsolution} I have used an extremely simple, yet numerically
precise way of solving the Gross-Pitaevskii equation in $1-$d by 
minimizing over a set of spline test functions. The program is made 
available on request. In $2-$d, I have used a finite-element approach,
which provides the most precise solution possible. This solution 
modifies by a few percent data which were kindly sent to me by 
F. Dalfovo.
\end{references}
\end{document}